\documentclass[reqno,draft]{amsart}

\theoremstyle{plain}
\newtheorem{lemma}{Lemma}
\newtheorem{theorem}[lemma]{Theorem}

\newtheorem{definition}[lemma]{Definition}
\theoremstyle{remark}
\newtheorem{remark}{Remark}

\newcommand*  {\cA}{{\mathcal A}}

\newcommand* {\cJ}{{\mathcal J}}
\newcommand*  {\cT}{{\mathcal T}}
\newcommand*  {\cV}{{\mathcal V}}

\newcommand*  {\bR} {{\bf R}}
\newcommand*  {\bZ} {{\bf Z}}

\renewcommand* {\sp}{{^\prime}}

\begin{document}

\title[ Three Dimensional Viscous Camassa--Holm Equations]%
{ The Three Dimensional Viscous Camassa--Holm Equations,
 and Their Relation to the Navier--Stokes Equations and
Turbulence Theory}

\date{Janaury 14, 2001}
\thanks{\textit{Journal of Dynamics and Differential Equations}, submitted.}

\author[C. Foias]{Ciprian Foias}
\address[ C. Foias]%
{ Departement of Mathematics\\ 
Texas A \& M University\\
College Station\\
TX 77843-3368, USA; also
Department of Mathematics \\
 Indiana University \\
 Bloomington, IN 47405, USA \\
and 1997-1998 Ulam Scholar at Los Alamos National Laboratory\\
 Center for Nonlinear Sciences, MS B258 \\
 Los Alamos, NM 87545, USA}
\email{foias@indiana.edu  \; foias@math.tamu.edu}

\author[D.D. Holm]{Darryl D. Holm}
\address[D.D. Holm]%
{Mathematical Modeling and Analysis, Theoretical Division, MS - B284 \\
Los Alamos National Laboratory \\
 Los Alamos, NM 87545, USA}
\email {dholm@lanl.gov}

\author[E.S. Titi]{Edriss S. Titi}
\address[E.S. Titi]%
{Department of Mathematics and
 Department of Mechanical and Aerospace Engineering \\
 University of California \\
 Irvine, CA 92697, USA\\
and 1997-1998 Orson Anderson Scholar at Los Alamos National Laboratory\\
 Institute of
Geophysics and Planetary Physics, IGPP, MS C305 \\
 Los Alamos, NM 87545 \\
USA}
\email{etiti@math.uci.edu}
\begin{abstract}
We show here the global, in time, regularity of the three dimensional
viscous Camassa-Holm (Navier--Stokes-alpha) equations. We also provide
estimates, in terms of the physical parameters of the equations, for the
Hausdorff and fractal dimensions of their global attractor. In analogy
with the Kolmogorov theory of turbulence, we define a small spatial
scale,  $\ell_{\epsilon}$, as the scale at which the balance occurs in
the mean rates of nonlinear transport of energy and viscous dissipation
of energy. Furthermore, we show that the number of degrees of freedom in
the long-time behavior of the solutions to these equations is bounded
from above by $\left(L/\ell_{\epsilon} \right)^3$, where $L$ is a
typical large spatial scale (e.g., the size of the domain). This
estimate suggests that the Landau--Lifshitz classical theory of
turbulence is suitable for interpreting the solutions of the
NS$-\alpha$ equations. Hence, one may consider these equations as a
closure model for the Reynolds averaged Navier--Stokes equations (NSE).
We study this approach, further, in other related papers.  Finally,
we discuss the relation of the NS$-\alpha$ model to the NSE by
proving a convergence theorem, that as the length scale $\alpha_1$
tends to zero a subsequence of solutions of the NS$-\alpha$ equations
converges to a weak solution of the three dimensional NSE.

\end{abstract}

\maketitle

\begin{center}
{\it  Dedicated to Professor Roger Temam on the
 Occasion of his $60^{th}$ Birthday}
\end{center}

\section{Introduction}

Proving global regularity for the  $3D$ Navier--Stokes equations (NSE)
is one of the most challenging outstanding problems in nonlinear
analysis.   The main difficulty in establishing this result lies in
controlling certain norms of vorticity.  More specifically, the
vorticity stretching term in the $3D$ vorticity equation forms the main
obstacle to achieving this control.

In this paper we consider a similar partial differential equation, the
so-called viscous Camassa--Holm, or Navier--Stokes-alpha (NS$-\alpha$)
equations.  The inviscid NS$-\alpha$  equations (Euler$-\alpha$)
were introduced in \cite{Holm1} as a natural mathematical
generalization of the integrable inviscid $1D$ Camassa--Holm equation
discovered in \cite{Camassa-Holm} through a variational
formulation.  Our studies in \cite{Chen1}--\cite{Chen3} indicated
that there is a connection between the solutions of the NS$-\alpha$ and
turbulence. Specifically, the explicit steady analytical solution of the
NS$-\alpha$ equations were found to compare successfully  with
empirical and numerical experimental data for mean velocity and
Reynolds stresses for turbulent flows in pipes and channels. These
comparisons led us to identify the NS$-\alpha$ equations with the
Reynolds averaged Navier--Stokes equations. These comparisons also led
us to suggest the NS$-\alpha$ equations could be used as a closure
model for the mean effects of subgrid excitations. Numerical tests that
tend to justify this intuition were reported in
\cite{Chen4}.

An alternative more ``physical'' derivation  for the inviscid
NS$-\alpha$  equations (Euler$-\alpha$), was introduced in
\cite{Holm2} and \cite{Holm3} (see also \cite{Chen2}). This alternative
derivation was based on substituting in Hamilton's principle the
decomposition of the Lagrangian fluid-parcel trajectory into its
mean and fluctuating components. This was followed by truncating a
Taylor series approximation and averaging at constant Lagrangian
coordinate, before taking variations. A variant of this approach was
also elaborated considerably in \cite{Marsden1}. See also
\cite{Marsden2} for the geometry and analysis of the Euler$-\alpha$
equations. For more information and a brief guide to the previous
literature specifically about the NS$-\alpha$ model, see paper
\cite{Foias-Holm-Titi2}. The latter paper also discusses  connections
to standard concepts and scaling laws in turbulence modeling, including
the relationship of the NS$-\alpha$ model to large eddy simulation
(LES) models. Results interpreting the NS$-\alpha$ model as an
extension of scale similarity LES models of turbulence are reported in
\cite{Dom-Holm}.

Equations similar to the NS$-\alpha$ equation, but
with different dissipative terms, were considered
previously in the theory of second grade
fluids~\cite{DF1974} and were treated recently in
the mathematical
literature~\cite{CV1996},~\cite{CV1997}. Second
grade fluid models are derived from continuum
mechanical principles of objectivity and material
frame indifference, after which thermodynamic
principles such as the Clausius-Duhem relation and
stability of stationary equilibrium states are
imposed that restrict the allowed values of the
parameters in these models. In contrast, as
mentioned earlier, the NS$-\alpha$ equation is
derived by applying asymptotic expansions,
Lagrangian means, and an assumption of isotropy of
fluctuations in Hamilton's principle for an ideal
incompressible fluid. Their different derivations
also provide the different interpretations of the
parameter $\alpha_1$, namely, as a flow regime
quantity for the NS$-\alpha$ equation, and
as a fixed material property for the second grade
fluid.

The aim of this paper is to establish the global  regularity of
solutions of the NS$-\alpha$, subject to periodic boundary
conditions.  We  also provide estimates of the fractal and Hausdorff
dimensions of their global  attractors.  In particular, we identify the
dimension of the attractor with the  number of degrees of freedom
governing the permanent regime of these equations and find  a remarkable
compatibility between these estimates and the number of degrees of
freedom in turbulence a la Landau and Lifshitz
\cite{Landau-Lifshitz}.  This leads us  to regard the NS$-\alpha$
equations as a suitable closure model for turbulence, thought of as an
averaged theory rather than an individual realization, cf.
\cite{Chen1}, \cite{Chen2}, \cite{Chen3}, \cite{ Holm2}
and \cite{Holm3}.
Finally, we relate the solutions of the viscous Camassa--Holm
(NS$-\alpha$) equations to those of the $3D$ NSE as the length scale
$\alpha_1$ tends to zero. Specifically, we prove that a subsequence of
solutions to the NS$-\alpha$ model converges as $\alpha_1\to0$ to a
weak solution of the $3D$ NSE.

\section{  Functional setting and preliminaries}
\label{s.preliminaries}
We consider the following viscous version of the three dimensional
Camassa--Holm equations in the periodic box $\Omega = [0,L]^3$:
\begin{subequations}
\label{CHeqn}
\begin{align}
\frac{\partial}{\partial t} (\alpha_0^2 u - \alpha_1^2 \Delta u) & -
\nu\Delta (\alpha_0^2 u - \alpha_1^2 \Delta u) - \nonumber \\
&- u \times (\nabla \times (\alpha_0^2 u - \alpha_1^2 \Delta u)) +
\frac{1}{ \rho_0}\nabla p = f \,, \\
\nabla\cdot u & = 0 \\
u(x,0) & = u^{in}(x) \, ,
\end{align}
\end{subequations}
where
${\displaystyle \frac{p}{\rho_0} = \frac{\pi}{\rho_0}}+
\alpha_0^2 |u|^2 - \alpha_1^2 (u \cdot \Delta u)
$
is the modified
pressure, while $\pi$ is the pressure, $\nu > 0$ is the constant viscosity
and $\rho_0 > 0$ is a constant density.  The function $f$ is a given body
forcing $\alpha_0 > 0$ and $\alpha_1 \geq 0$ are scale parameters.  Notice
$\alpha_0$ is dimensionless while $\alpha_1$ has units of length.
Also observe that at the limit $\alpha_0 = 1, \alpha_1 = 0$ we obtain
the three dimensional Navier--Stokes equations with periodic boundary
conditions.

For simplicity we will assume the forcing term to be time independent, i.e.
$f(x,t)\equiv f(x)$.

{F}rom~(\ref{CHeqn}) one can easily see, after
integration by parts, that

$$\frac{d}{dt}  \int_\Omega (\alpha_0^2 u - \alpha_1^2 \Delta u) dx =
\int_\Omega fdx\;.$$

On the other hand, because of the spatial periodicity of the solution,
we have $\int_\Omega \Delta udx = 0$.  As a result, we have
$\frac{d}{dt} \int_\Omega udx = \int_\Omega f dx$; that is, the mean of the
solution is invariant provided the mean of the forcing term is zero.
In this paper we will consider forcing terms and initial values with
spatial means that are zero; i.e., we will assume
$\int_\Omega u^{in}dx = \int fdx = 0$ and hence $\int_\Omega udx = 0$.

Next, let us introduce some notation and background.
\medskip

\begin{enumerate}
\item[(i)] Let $X$ be a linear subspace of integrable functions defined on
the domain $\Omega$, we denote

$$\dot X := \{\varphi\in X : \int_\Omega \varphi(x) dx = 0\}\;.$$

\item[(ii)] We denote $ \cV = \{\varphi:\varphi$ is a vector valued
trigonometric polynomial defined on $\Omega$, such that
$\nabla\cdot\varphi= 0$ and $\int_\Omega \varphi(x)dx = 0\}$, and let
$H$ and $V$ be the closures of $\cV$ in $L^2(\Omega)^3$ and in $H^1
(\Omega)^3$ respectively; observe that $H^\perp$, the orthogonal complement
 of $H$ in
$L^2(\Omega)^3$ is $\{\nabla p: \quad p\in H^1(\Omega)\}$
(cf.  \cite{Constantin-Foias-book} or \cite{Temam-book1}).

\item[(iii)] We denote $P_\sigma : \dot L^2 (\Omega)^3\to H$ the
$L^2$ orthogonal projection, usually referred as Leray projector,
and by $A = - P_\sigma \Delta$ the Stokes operator with domain
$D(A) = (H^2(\Omega))^3 \cap V$.  Notice that in the case of periodic boundary
condition $A = -\Delta\big|_{D(A)}$ is a self-adjoint positive
operator with compact inverse.  Hence the space $H$ has an
orthonormal basis $\{w_j\}^\infty _{j=1}$ of eigenfunctions of $A$,
i.e. $Aw_j = \lambda_j w_j$, with $0 < \lambda _1 \leq \lambda _2
\ldots \leq \lambda _j \to \infty $; in fact these eigenvalues have
the form $|k|^2 \frac{4\pi^2}{L^2}$ with $k\in \bZ^3\backslash\{0\}$.

\item[(iv)] We denote $(\cdot,\cdot)$ the $L^2$-inner product and
by $|\cdot|$ the corresponding $L^2$-norm.  By virtue of Poincar\' e
inequality one can show that there is a constant $c>0$ such that
$$c|Aw| \leq \|w\|_{H^2} \leq  c^{-1} |Aw|\ \hbox{for every}\ w\in D(A)$$
and that
$$c|A^{1/2}w| \leq \|w\|_{H^1} \leq c^{-1}|A^{1/2}w|\ \hbox{for every}\
w\in V\;.$$
Moreover, one can show that  $V = D(A^{1/2})$, (cf.
\cite{Constantin-Foias-book} , and \cite{Temam-book1}). We denote
$((\cdot, \cdot )) = (A^{1/2} \cdot, A^{1/2} \cdot)$  and $\| \cdot\|=
|A^{1/2} \cdot|$ the inner product and norm on $V$, respectively. Notice that,
based on the above, the inner product $((\cdot, \cdot )) $,  restricted to
$V$, is equivalent to the $H^1$  inner product %
\begin{equation}
\label{inner}
[u,v] = \alpha_0^2 (u,v) + \alpha_1^2((u,v)) \quad {\hbox{for}} \quad
u,v \in V \;,
\end{equation}
provided $\alpha_1 >0$.

Hereafter $c$ will denote a generic scale invariant  positive
constant which is independent of the physical parameters in
the equation.

\item[(v)] Following the notation for the Navier--Stokes equations we
denote $B(u,v)= P_\sigma  [(u\cdot \nabla) v]$, and we set
$B(v)u = B(u,v)$ for every $u,v \in V$. Notice that
\begin{equation}
\label{BI1}
(B(u,v), w) = -(B(u,w), v) \, \, {\mbox{ for every}} \, \,  u,v,w \in V
\end{equation}
We also denote $\tilde B (u,v) = -P_\sigma  (u\times (\nabla
\times v))$ for every $u,v\in V$.  Using the identity
$$
(b\cdot\nabla)a + \sum^3_{j=1} a_j \nabla b_j = -b\times (\nabla
\times a) + \nabla (a\cdot b) \, ,
$$
one can easily show that
\begin{equation}
\label{BI2}
\begin{align}
(\tilde B (u,v),w) & =  (B (u,v),w) - (B (w,v),u) \nonumber \\
& = (B(v)u -B^*(v)u,w) \, ,
\end{align}
\end{equation}
for every $u,v,w\in V\;$.  As a result we have
\begin{equation}
\label{BI3}
\tilde B (u,v) = (B(v) -B^*(v))u \, \, {\mbox{ for every}} \, \, u,v \in V \, .
\end{equation}

\end{enumerate}

In the next Lemma, we show that the bilinear operator $\tilde B$ can
be extended continuously to a larger class of functions.

\begin{lemma} \label{Properties}
\begin{enumerate}
\item[(i)] The operator $A$ can be extended continuously to be defined on
$V= D(A^{1/2})$ with values in $V\sp \subset H^{-1}$ such that
$$
\langle Au,v\rangle_{V\sp} = (A^{1/2}u, A^{1/2}v) = \int_\Omega
(\nabla u: \nabla v)dx
$$
for every $u,v\in V$.

\item[(ii)] Similarly, the operator $A^2$ can be extended continuously to be
defined on $D(A)$ with values in $D(A)\sp$ such that
$$
\langle A^2u,v\rangle_{D(A)\sp} = (Au,Av)\;,\; \hbox{for every}\
u,v\in D(A)\;.
$$

\item[(iii)] The operator $\tilde B$ can be extended
continuously from $V\times V$ with values in $V\sp$, and in particular
it satisfies
\begin{eqnarray*}
|\langle\tilde B (u,v),w\rangle_{V\sp}| &\leq& c|u|^{1/2}
\|u\|^{1/2} \|v\| \|w\|  \\
|\langle\tilde B (u,v),w\rangle_{V\sp}| &\leq& c\|u\|
\|v\| |w|^{1/2} \|w\|^{1/2}\; ,
\end{eqnarray*}
for every $u,v,w\in V$.    Moreover,
$$
\langle\tilde B (u,v),w\rangle_{V\sp} = -\langle\tilde
B(w,v),u\rangle_{V\sp}\;,\ \hbox{for every}\ u,v,w\in V\; ,
$$
and in particular,
$$
\langle\tilde B (u,v),u\rangle_{V\sp} \equiv 0\ \hbox{for every}\
u,v\in V\;.
$$

\item[(iv)] Furthermore, we have
$$
\left|\langle \tilde B (u,v),w\rangle_{D(A)\sp}\right| \leq c|u|
\|v\| \|w\|^{1/2} |Aw|^{1/2}\;,
$$
for every $u\in H, ~ v\in V\ \hbox{and}\ w\in D(A)$,
and by symmetry we have
$$
\left | ( \tilde B (u,v),w) \right| \leq
c\|u\|^{1/2} |Au|^{1/2} \|v\| |w|
$$
for every $ u\in D(A), v\in V\ \hbox{and}\ w\in H\;. $

\item[(v)] Also,
$$
\left|\langle \tilde B (u,v),w\rangle_{D(A)\sp}\right| \leq
c \left (|u|^{1/2} \|u\|^{1/2} |v| |Aw| + |v| \|u\| \|w\|^{1/2}
|Aw|^{1/2} \right )\;,
$$
for every $ u\in V , v\in H, w\in D(A)\;.$

\item[(vi)] In addition,
$$
\left|\langle \tilde B (u,v),w\rangle_{V\sp}\right| \leq
c \left (\|u\|^{1/2} \|Au\|^{1/2} |v| \|w\| + |Au| |v| |w|^{1/2}
\|w\|^{1/2}\right) \;,
$$
 for every $ u\in D(A), v\in H, w\in V\;.$
\end{enumerate}
\end{lemma}

\begin{proof}
 The proof of (i) can be found in \cite{Constantin-Foias-book} or in
\cite{Temam-book1}.  The proof of (ii) is a straight forward extension of
that of (i).

To prove (iii), let us first consider the case when $u,v,w\in \cV$.
Then we have
\begin{eqnarray*}
|\langle \tilde B (u,v),w\rangle_{V\sp}| &=&
|\int_\Omega u\times(\nabla\times v)\cdot wdx | \\
&\leq & c\|u\|_{L^3} \|\nabla v\|_{L^2} \|w\|_{L^6}\;.
\end{eqnarray*}
Recall the following Sobolev inequalities in $\bR^3$
\begin{subequations}
\label
{Sobolev}
\begin{gather}
\begin{align}
\|\varphi\|_{L^4} &\leq  c\|\varphi\|^{1/4}_{L^2}
\|\varphi\|^{3/4}_{H^{1}}\; , \\
\|\varphi\|_{L^3} &\leq  c\|\varphi\|^{1/2}_{L^2}
\|\varphi\|^{1/2}_{H^{1}}\; , \; \hbox{and}\\
\|\varphi\|_{L^6} &\leq  \|\varphi\|_{H^1}\;,\; \hbox{for every}\
\varphi\in H^1(\Omega)\;.
\end{align}
\end{gather}
\end{subequations}
Then by the above inequalities we have:
$$
|\langle \tilde B (u,v),w\rangle_{V\sp}| \leq c|u|^{1/2}
\|u\|^{1/2} \|v\| \|w\|\;.
$$
Moreover, it is clear that for $u,v,w\in \cV$
$$
\langle \tilde B (u,v),w\rangle_{V\sp} = -\langle \tilde
B(w,v),u\rangle_{V\sp}\;.
$$
Since $\cV$ is dense in $V$ we conclude the proof of (iii).

Let us now prove (iv).  Again we consider first the case where
$u,v,w\in \cV$
\begin{equation}
\begin{align}
|\langle \tilde B (u,v),w\rangle_{D(A)\sp}| &=
|\int_\Omega [u\times (\nabla\times v)]\cdot wdx| \nonumber \\
&\leq c\|u\|_{L^2} \|\nabla v\|_{L^2} \|w\|_{L^\infty }\;. \nonumber
\end{align}
\end{equation}
Recall Agmon's inequality in $\bR^3$:
\begin{equation}
\label{Agmon}
\|\varphi\|_{L^\infty } \leq c\|\varphi\|_{H^1}^{1/2}
\|\varphi\|_{H^2}^{1/2}\;.
\end{equation}
The above gives
$$
|\langle \tilde B (u,v),w\rangle_{D(A)\sp}| \leq
c |u|~ \|v\|~ \|w\|^{1/2} \|Aw\|^{1/2}\;.
$$

To prove ( v) we again take $u,v,w\in \cV$ and we
use~(\ref{BI2}) to find
\begin{eqnarray*}
|\langle \tilde B (u,v),w\rangle_{D(A)\sp}| &\leq &
|\int_\Omega ((u\cdot\nabla)v)\cdot wdx|
+ |\int_\Omega ((w\cdot\nabla)u)\cdot
vdx|\\
&\leq & |\int_\Omega ((u\cdot\nabla)w\cdot v)dx| +
\|v\|_{L^2} \|\nabla u\|_{L^2} \|w\|_{L^\infty }\\
&\leq & c\|u\|_{L^3} \|\nabla w\|_{L^6} |v| + c|v|~ \|u\| ~
\|w\|_{L^\infty}\;.
\end{eqnarray*}
By~(\ref{Sobolev}b-c) and~(\ref{Agmon}) inequalities we finish our proof.

The proof of (vi) is similar to (v).  From~(\ref{BI2}) we have
\begin{eqnarray*}
|\langle \tilde B (u,v),w\rangle_{V\sp}| &\leq &
|\int_\Omega ((u\cdot\nabla)v) wdx|
+ |\int_\Omega ((w\cdot\nabla)u)\cdot vdx| \\
&\leq &|\int_\Omega ((u\cdot\nabla)w)\cdot vdx| +
c\|w\|_{L^3} \|\nabla u\|_{L^6} |v| \\
&\leq & c (\|u\|_{L^\infty} \|w\| ~|v| + \|w\|_{L^3} \|\nabla u\|_{L^6}
|v|)\;.
\end{eqnarray*}
By~(\ref{Sobolev}a) and~(\ref{Agmon}) inequalities
we finish our proof.

\end{proof}

We apply  $P_\sigma $ to~(\ref {CHeqn}) and use the above
notation to obtain the equivalent system of equations
\begin{subequations}
\label{The-eqn1}
\begin{align}
\frac{d}{ dt} (\alpha_0^2 u + \alpha_1^2 Au) &+ \nu A(\alpha_0^2 +
\alpha_1^2 A)u +  \nonumber \\
&+ \tilde B(u,\alpha_0^2 u + \alpha_1^2 Au) = P_\sigma  f\; ,\\
 u(0)  & = u^{in}\; .
\end{align}
\end{subequations}
Alternatively, if we denote
\begin{equation}
\label{v}
v= \alpha_0^2 u + \alpha_1^2 Au
\end{equation}
the system~(\ref{The-eqn1}) can be written as
\begin{subequations}
\label{The-eqn2}
\begin{gather}
\begin{align}
\frac{dv}{ dt} + \nu A v &+ B(v)u -B^*(v)u = P_\sigma  f\; ,\\
 u(0)  & = u^{in}\; .
\end{align}
\end{gather}
\end{subequations}
We will assume that $P_\sigma f = f$, otherwise we add the gradient
part of $f$ to the modified pressure and rename $P_\sigma f$ by $f$.
\bigskip

\begin{definition}[Regular Solution]  Let $f\in H$, and let
$T > 0$.  A function $u\in C([0,T);V) \cap L^2 ([0,T); D(A))$
with $\frac{du}{dt} \in L^2 ([0,T);H)$ is said to be a regular solution to
(\ref {The-eqn1}) in the interval $[0,T)$ if it satisfies
\begin{equation}
\label{weak1}
\begin{align}
\langle \frac{d}{dt} (\alpha_0^2 u + \alpha_1^2 Au),
w\rangle_{D(A)\sp}
+ \nu \langle A(\alpha_0^2 u + \alpha_1^2 Au), & w\rangle_{D(A)\sp}
\nonumber \\
+ \langle \tilde B(u,\alpha_0^2 u + \alpha_1^2
Au),w\rangle_{D(A)\sp} = & (f,w) \; ,
\end{align}
\end{equation}
for every $w\in D(A)$ and for almost every $t\in [0,T)$.  Moreover,
$u(0) = u^{in}$ in $V$.
Here, the equation~(\ref{weak1}) is
understood in the following sense:
\smallskip

\noindent For every $t_0, t\in [0,T)$ we have

\begin{equation}
\label{weak2}
\begin{align}
(\alpha_0^2 u(t) + \alpha_1^2 Au(t),w) + \nu \int^t_{t_0} (\alpha_0^2
u(s) + \alpha_1^2 Au(s),w)ds + & \nonumber \\
+ \int^t_{t_0} \langle \tilde B (u(s), \alpha_0^2 u(s) + \alpha_1^2
Au(s)), w\rangle_{D(A)\sp} ds & = \int^t_{t_0} (f,w) ds\; .
\end{align}
\end{equation}
\end{definition}

\section{Global existence and uniqueness}
\label{existence-uniqueness}

In this section we prove global existence and uniqueness of regular
solutions to equation (\ref{The-eqn1}).
\medskip

\begin{theorem} [Global existence and uniqueness]
\label{Global}
 Let $f\in H$ and $u^{in}\in V$.  Then for any $T > 0$ the equation
(\ref{The-eqn1}) has  a unique
regular solution $u$ on $[0,T)$.  Moreover, this solution satisfies:
\begin{enumerate}
\item[(i)]  $u\in L^\infty _{\rm loc} ((0,T]; H^3(\Omega))$.

\item[(ii)]  There are constants $R_k $, for $k  = 0,1,2,3$,
which depend only on $\nu ,\alpha_0,\alpha_1$ and $f$, but not on
$u^{in}$, such that
$$
\limsup_{t\to\infty } \left(\alpha_0^2 |A^{\frac{k}{2}}u|^2 + \alpha_1^2
|A^{\frac{k +1}{ 2}}u|^2\right) = R^2_k \; ,
$$
for $k =0, 1,2,3$. In particular, we have
\begin{equation}
\label{R}
R_0^2 = \frac{1}{\nu \lambda _1} \min\left\{\frac{|A^{-1/2}f|^2}{\nu
\alpha_0^2}\;,\; \frac{|A^{-1/2}f|^2}{\nu \alpha_1^2}\right\} \leq
\min\left\{ \frac{|f|^2}{\nu ^2\lambda_1 ^2\alpha_0^2}\;,\;
\frac{|f|^2}{\nu ^2\lambda_1 ^3\alpha_1^2}\right\}\;,
\end{equation}
that is:
$$
R^2_0 \leq \frac{G^2 \nu^2}{ \lambda_1^{1/2}}
\min\left\{\frac{1}{\alpha_0^2}\;,\; \frac{1}{\alpha_1^2\lambda_1}\right\}
 = \frac{G^2\nu^2}{\gamma \lambda_1^{1/2}}\; ,
$$
where $G = \frac{|f|}{\nu ^2\lambda_1^{3/4}}$ is the Grashoff number,
and $\gamma^{-1} = \min\left\{\frac{1}{\alpha_0^2}\;,\;
\frac{1}{ \alpha_1^2\lambda_1}\right\}\;.$
Furthermore,
\begin{equation}
\label{Mean-R}
\limsup_{T\to\infty } \frac{\nu}{ T} \int^{t+T}_t (\alpha_0^2
\|u(s)\|^2 + \alpha_1^2 |Au(s)|^2)ds \leq \nu \lambda _1 R_0^2 \leq
\frac{G^2 \nu ^3\lambda_1^{1/2}}{ \gamma }\; ,
\end{equation}
for all $t \geq 0$.
\end{enumerate}
\end{theorem}

\begin{proof}

  We use the Galerkin procedure to prove global
existence and to establish the necessary a priori estimates.

Let $\{w_j\}^\infty _{j=1}$ be an orthonormal basis of $H$ consisting
of eigenfunctions of the operator $A$.  Denote $H_m =\;{\rm span}
\{w_1,\ldots,w_m\}$ and let $P_m$ be the $L^2$-orthogonal projection
from $H$ onto $H_m$.  The Galerkin procedure for the equation
(\ref{The-eqn1})  is the ordinary differential system.
\begin{subequations}
\label{Gal}
\begin{align}
\frac{d}{ dt} (\alpha_0^2 u_m + \alpha_1^2 Au_m) &+
\nu  A(\alpha_0^2 u_m + \alpha_1^2 Au_m) + \nonumber \\
& + P_m \tilde B (u_m, \alpha_0^2 u_m + \alpha_1^2 Au_m) = P_m f \;,\\
& u_m(0) = P_m u^{in}\;.
\end{align}
\end{subequations}
Since the nonlinear term is quadratic in $u_m$, then by the classical
theory of ordinary differential equations, the system~(\ref{Gal})
has a unique solution for a short interval of time
$(-\tau_m, T_m)$.  Our goal is to show that the solutions of~(\ref{Gal})
remains finite for all positive times which
implies that $T_m = \infty $.

\bigskip
\noindent{\it \underbar{$H^1$-estimates}}

We take the inner product of (\ref{Gal}) with $u_m$ and use~(\ref{BI2})
to obtain
$$
\frac{1}{2} \frac{d}{ dt} (\alpha_0^2|u_m|^2 + \alpha_1^2 \|u_m\|^2) +
\nu (\alpha_0^2 \|u_m\|^2 + \alpha_1^2 |Au_m|^2) = (P_mf,u_m)\;.
$$
Notice that
\[
|(P_mf,u_m)| \leq \left \{ \begin{array}{l}
|A^{-1}f| |Au_m| \\
|A^{-1/2}f| \|u_m\| \; , \end{array}\right.
\]
and by Young's inequality we have
\[
|(P_mf,u_m)| \leq  \left \{ \begin{array}{l}
{\displaystyle \frac{|A^{-1}f|^2}{ 2\nu \alpha_1^2}} +
\frac{\nu}{2}\alpha_1^2 |Au_m|^2 \\
{\displaystyle \frac{|A^{-1/2}f|^2}{2\nu \alpha_0^2}}+
\frac{\nu}{ 2}\alpha_0^2 \|u_m\|^2\;. \end{array} \right.
\]
Denoting by $K_1 = \min\{{\displaystyle
\frac{|A^{-1/2}f|^2}{ \nu  \alpha_0^2},
 \frac{|A^{-1}f|^2}{ \nu \alpha_1^2}}\}$,
from the above inequalities we get: 
\begin{equation}
\label{H1}
\frac{d}{dt} (\alpha_0^2 |u_m|^2 + \alpha_1^2 \|u_m\|^2) +
\nu  (\alpha_0^2\|u_m\|^2 + \alpha_1^2 |Au_m|^2) \leq K_1\;.
\end{equation}
By Poincar\'{e}'s inequality we obtain
$$
\frac{d}{ dt} (\alpha_0^2 |u_m|^2 + \alpha_1^2 \|u_m\|^2) +
\nu \lambda_1 (\alpha_0^2 |u_m|^2 + \alpha_1^2 \|u_m\|^2) \leq K_1
$$
and then by Gronwall's inequality we reach
\begin{eqnarray*}
\alpha_0^2 |u_m(t)|^2 + \alpha_1^2 \|u_m(t)\|^2 &\leq &
e^{-\nu \lambda_1 t}  (\alpha_0^2 |u_m(0)|^2 + \alpha_1^2\|u_m(0)\|^2) +
\nonumber \\
&&+ \frac{K_1}{\nu \lambda_1} (1 - e^{-\nu \lambda_1t}) \; .
\end{eqnarray*}
That is
\begin{equation}
\label{H1-bound}
\alpha_0^2 |u_m(t)|^2 + \alpha_1^2 \|u_m(t)\|^2 \leq
k_1 := \alpha_0^2 |u^{in}|^2 + \alpha_1^2 \|u^{in}\|^2 +
\frac{K_1}{\nu \lambda_1}\; .
\end{equation}

\smallskip

\noindent {\it \underbar{$H^2$-estimates}}

Integrating~(\ref{H1}) over the interval $(t,t+\tau )$
\begin{equation}
\label{H2-ave}
\begin{align}
\nu  \int^{t+\tau }_t (\alpha_0^2 \|u_m(s)\|^2 + \alpha_1^2
|Au_m(s)|^2)ds & \leq \tau  K_1
+ (\alpha_0^2 |u_m(t)|^2 + \alpha_1^2 \|u_m(t)\|^2) \nonumber \\
 &\leq \tau K_1  + k_1 =: \bar k_2(\tau) \;.
\end{align}
\end{equation}
Now, take the inner product of~(\ref {Gal}) with $Au_m$ to
obtain
\begin{equation}
\begin{align}
\frac{1}{ 2} \frac{d}{ dt} (\alpha_0^2\|u_m\|^2 + \alpha_1^2 |Au_m|^2) & +
\nu (\alpha^2_0 |Au_m|^2 + \alpha_1^2|A^{3/2}u_m|^2) + \nonumber \\
&+ (\tilde B(u_m, \alpha^2_0 u_m + \alpha_1^2 Au_m),Au_m) =
(P_mf, Au_m)\;. \nonumber
\end{align}
\end{equation}
Notice that
\[
|(P_mf, Au_m)| \leq \left\{ \begin{array}{l}
|A^{-1/2}f| |A^{3/2}u_m|\\
|f| |Au_m|  \end{array} \right.
\leq \left\{ \begin {array}{l}
\frac{|A^{-1/2}f|^2}{ \nu \alpha_1^2} + \frac{\nu}{ 4} \alpha_1^2
|A^{3/2} u_m|^2 \\
\frac{|f|^2}{\nu \alpha_0^2} + \frac{\nu}{ 4} \alpha_0^2
|Au_m|^2\;. \end{array}\right.
\]
We denote $K_2 = \min\left\{ \frac{|A^{-1/2}f|^2}{ \nu \alpha_1^2},
\frac{|f|^2}{ \nu \alpha_0^2}\right\}$.  Then we have
\begin{equation}
\begin{align}
\frac{1}{ 2} \frac{d}{dt} (\alpha_0^2 \|u_m\|^2 + \alpha_1^2 |Au_m|^2)&+
\frac{3\nu}{ 4} (\alpha_0^2 |Au_m|^2 + \alpha_1^2 |A^{3/4}u_m|^2) +
\nonumber \\
& + (\tilde B(u_m, \alpha_0^2 u_m + \alpha_1^2 Au_m),Au_m) \leq K_2\;.
\nonumber
\end{align}
\end{equation}
We use part (iii) of Lemma~(\ref  {Properties}) to obtain
\begin{equation}
\begin{align}
\frac{1}{ 2} \frac{d}{ dt} (\alpha_0^2 & \|u_m\|^2 + \alpha_1^2 |Au_m|^2) +
\frac{3}{ 4}\nu  (\alpha_0^2 |Au_m|^2 + \alpha_1^2 |A^{3/2}u_m|^2)
\leq  \nonumber \\
&\leq c\|u_m\|(\alpha_0^2 \|u_m\| + \alpha_1^2 |A^{3/2}u_m|)
|Au_m|^{1/2} |A^{3/2}u_m|^{1/2} + K_2 \nonumber \\
&\leq c \|u_m\| (\alpha_0^2 \lambda_1^{-1} + \alpha_1^2) |A^{3/2} u_m|^{3/2}
|Au_m|^{1/2} + K_2 \; .
\nonumber
\end{align}
\end{equation}
By Young's inequality we have
\begin{equation}
\label{H2}
\begin{align}
\frac{1}{ 2} \frac{d}{ dt} (\alpha_0^2 \|u_m\|^2 & + \alpha_1^2 |Au_m|^2) +
\frac{\nu }{ 2} (\alpha_0^2 |Au_m|^2 + \alpha_1^2 |A^{3/2} u_m|^2)
 \nonumber \\
& \leq c\|u_m\|^4 (\alpha_0^2 \lambda_1^{-1} + \alpha_1^2)^4
(\nu \alpha_1^2)^{-3} |Au_m|^2 + K_2 \;.
\end{align}
\end{equation}
We integrate the above equation over $(s,t)$ and use~(\ref {H1-bound}) and
(\ref{H2-ave})
to obtain:
\begin{equation}
\begin{align}
\alpha_0^2 \|u_m(t)\|^2 &+ \alpha_1^2 |Au_m(t)|^2 \leq
\alpha_0^2\|u_m(s)\|^2 + \alpha_1^2 |Au_m(s)|^2 +   \nonumber \\
&+ 2(t-s) K_2 +
\frac{2c k_12}{(\nu \alpha_1^2)^4 \alpha_1^4}(\alpha_0^2 \lambda_1^{-1} +
\alpha_1^2)^4 [(t-s) K_1 +k_1] \nonumber
\end{align}
\end{equation}

Now, we integrate with respect to $s$ over $(0,t)$ and use~(\ref{H2-ave})
 to get
\begin{equation}
\label{tH2-bound}
\begin{align}
t(\alpha_0^2 \|u_m(t)\|^2 &+ \alpha_1^2 |Au_m(t)|^2) \leq
\frac{1}{\nu} (tK_1+ k_1) +  t^2 K_2 +  \nonumber \\
&+ \frac{2c k_1^2}{(\nu \alpha_1^2)^4 \alpha_1^4}(\alpha_0^2 \lambda_1^{-1} +
\alpha_1^2)^4 [\frac{t^2 K_1}{2} +t k_1] \;,
\end{align}
\end{equation}
for all $ t \ge 0$.

For $t \ge \frac{1}{\nu \lambda_1}$  we integrate with respect to
$s$ over the interval $(t- \frac{1}{\nu \lambda_1}, t)$
\begin{equation}
\label{H2-large-t-bound}
\begin{align}
\frac{1}{\nu \lambda_1}(\alpha_0^2 \|u_m(t)\|^2 &+
\alpha_1^2 |Au_m(t)|^2) \leq
\frac{1}{\nu} (\frac{1}{\nu \lambda_1}K_1+ k_1) +
(\frac{1}{\nu \lambda_1})^2 K_2 +  \nonumber \\
&+ \frac{2c k_1^2}{(\nu \alpha_1^2)^4 \alpha_1^4}(\alpha_0^2 \lambda_1^{-1} +
\alpha_1^2)^4 \left[(\frac{1}{\nu \lambda_1})^2\frac{ K_1}{2} +
\frac{k_1}{\nu \lambda_1} \right] \;.
\end{align}
\end{equation}
{F}rom~(\ref {tH2-bound}) and~(\ref{H2-large-t-bound}) we conclude:
\begin{equation}
\label{H2-bound}
\alpha_0^2 \|u_m(t)\|^2 +
\alpha_1^2 |Au_m(t)|^2  \leq k_2(t)
\end{equation}
for all $t > 0$, where $k_2(t)$ enjoys the following properties:
\begin{enumerate}
\item[(i)]  $k_2(t)$ is finite for all $t > 0$\;.

\item[(ii)]  $k_2(t)$ is independent of $m$\;.

\item[(iii)] If $u^{in}\in V$, but $u^{in}\notin D(A)$, then $k_2(t)$
depends on $\nu , f, \alpha_0$ and $\alpha _1$.  Moreover, in this
case $\lim_{t\to 0^+} k_2(t) = \infty $\;.
\item[(iv)] $\limsup_{t\to\infty } k_2(t) = R^2_2 < \infty$.
\end{enumerate}
\smallskip

Returning to~(\ref{H2}) and integrating over the interval
$(t,t+\tau)$, for $t > 0$ and $\tau\geq 0$ and using~(\ref {H2-bound})
we get
\begin{equation}
\label{H3-ave}
\int^{t+\tau}_t (\alpha_0^2 |Au_m(s)|^2 + \alpha_1^2 |A^{3/2}
u_m(s)|^2) ds \leq \bar k_3 (t,\tau)
\end{equation}
where $\bar k_3 (t,\tau)$ as a function of $t$ satisfies properties
(i)-(iii) as $k_2(t)$ above.  Also, there exists $T_1$ large enough,
 depends on $(\alpha_0^2 |u^{in}|^2 + \alpha_1^2 \|u^{in}\|^2)$,
but independent of $m$, such that
$$
\frac{1}{ t} \int^t_0 (\alpha_0^2 |Au_m(s)|^2 + \alpha_1^2 |A^{3/2}
u_m(s)|^2) ds \leq 2R^2_2\ \quad \hbox{for all}\ t>T_1\;.
$$

\bigskip

\noindent {\it \underbar{$H^3$ estimate}} (via the vorticity)

Let us denote $v_m = \alpha _0 u_m + \alpha _1 Au_m$ and
$q_m = \nabla \times v_m$. The Galerkin system~(\ref {Gal})
is equivalent to
$$
\frac{dv_m}{ dt} + \nu Av_m - P_m(u_m \times q_m) = P_mf.
$$
Let us take {\bf Curl} of the above equation, keeping in mind
that  we have
periodic boundary conditions, to obtain
$$
\frac {dq_m}{ dt} + \nu  Aq_m - \nabla\times(P_m(u_m \times q_m)) =
\nabla \times P_mf \,.
$$
Notice that $ \nabla \cdot q_m = 0$ and that $P_mq_m = q_m$.
Let us take the
inner product of the above equation with $q_m$
$$
\frac{1}{ 2} \frac{d}{ dt} |q_m|^2 + \nu \|q_m\|^2 - (\nabla\times
(P_m(u_m \times q_m)),q_m) = (\nabla\times P_mf,q_m)\;.
$$
We use the identity
\begin{equation}
\label{curl-I}
 \int_\Omega(\nabla\times \phi)\cdot \psi dx =
\int_\Omega \phi\cdot (\nabla\times \psi)dx,
\end{equation}
to reach
$$
\frac{1}{2} \frac{d}{ dt}|q_m|^2 + \nu \|q_m\|^2 - (P_m(u_m
\times q_m),\nabla \times q_m) = (P_mf,\nabla \times q_m)\;.
$$
Notice that $P_m (\nabla \times q_m) = \nabla \times q_m$, therefore
\[
\frac{1}{ 2} \frac{d}{ dt} |q_m|^2 + \nu \|q_m\|^2 =
(u_m \times q_m,\nabla\times q_m) + (f,\nabla \times q_m) \;,
\]
and upon applying~(\ref{curl-I})
\[
\frac{1}{2} \frac{d}{ dt} |q_m|^2 + \nu \|q_m\|^2 =
(\nabla\times(u_m \times q_m),q_m) + ( f,\nabla\times q_m)\;.
\]

For every divergence-free function $\phi$, and for every $\psi$ 
 we have the identity
$$
\nabla \times (\phi \times \psi) = -(\phi\cdot\nabla)\psi +
(\psi\cdot\nabla)\phi\;.
$$
As a result, we have
$$
\frac{1}{2} \frac{d}{ dt} |q_m|^2 + \nu \|q_m\|^2= -((u_m\cdot
\nabla) q_m,q_m) + (q_m\cdot\nabla u_m,q_m) + (f, \nabla\times  q_m)\;.
$$
Thanks to the identity~(\ref{BI1}) we have $((u_m\cdot\nabla)q_m,q_m) = 0$.
 Now, we estimate the
right hand side of the above to get:
$$
\frac{1}{2} \frac{d}{ dt} |q_m|^2 + \nu \|q_m\|^2 \leq
c\|q_m\|_{L^4} \|u_m\| + |f| \|q_m\;.$$
We use the Sobolev inequality~(\ref{Sobolev}a)
  and Young's inequality to find 
$$
\frac{1}{2} \frac{d}{ dt} |q_m|^2 + \nu \|q_m\|^2 \leq c\|q_m\|^{3/4}
|q_m|^{1/4} \|u_m\| + \frac{1}{ \nu } |f|^2 +\frac{\nu }{ 4}
\|q_m\|^2\;,
$$
and we use Young's inequality again to obtain
$$\frac{1}{2} \frac{d}{ dt} |q_m|^2  + \frac {\nu}{ 2} \|q_m\|^2 \leq
\frac {c}{\nu ^3} |q_m|^2  \|u_m\|^4 + \frac{1}{ \nu } |f|^2 \;.
$$
Let us denote $z_m(t) = \nu^2 \lambda_1^{1/2} + |q_m(t)|^2$, then
$$
\frac{dz_m}{ dt} \leq z_m(t) \left (\frac {c\|u_m(t)\|^4 }{ \nu ^3} +
\frac {|f|^2 }{ \nu^3 \lambda_1^{1/2}}\right )\;.
$$
We use~(\ref  {H1-bound}) to obtain
$$
z_m(t) \leq z_m(s) e^{\displaystyle \int_0^t (\frac{c k_1^2 }{ \nu^3
\alpha_1^4} +
\frac {|f|^2}{ \nu ^3\lambda_1^{1/2}})d\tau}
$$
for every $0,s\leq t$.  From the definition of $z_m$ we observe
$$
z_m(s) \leq c(\alpha_0^2 |Au_m(s)|^2 + \alpha_1^2 |A^{3/2}
u_m(s)|^2 +  \nu ^2\lambda _1^{1/2}).
$$
Now, we integrate with respect to $s$ over $(\frac{t}{2},t)$ and
use~(\ref {H3-ave}) to get
\begin{equation}
\label{H3-bound}
 z_m(t) \leq \left [ \frac {2}{ t} \bar k_3 (\frac{t}{ 2},\frac{t}{ 2}) +
\nu ^2\lambda _1^{1/2} \right ]
e^{\displaystyle \int_0^t (\frac{c k_1^2 }{ \nu^3 \alpha_1^4} +
\frac {|f|^2}{ \nu ^3\lambda_1^{1/2}})d\tau}
  =: k_3(t)\;.
\end{equation}
Here again $k_3(t)$ enjoys the properties (i)-(iii) of $k_2(t)$,
mentioned above.
\bigskip

\begin{remark} Notice that by establishing the estimate~(\ref{H3-bound})
for $|q_m|$ one indeed is providing an upper bound for the $H^3$-norm  of
$u_m$.  Similar estimates for the $H^3$-norm of $u_m$  can be
also obtained by considering first the Galerkin system~(\ref{Gal})
$$
\frac{dv_m}{dt} + \nu  Av_m + P_m\tilde B (u_m,v_m) =\ P_mf\;,
$$
taking the inner product with  $Av_m$, and then following  a
sequence of inequalities and estimates to achieve   an upper
bound for $\|v_m\|$.
\end{remark}

Let us now summarize our estimates. For any $T>0$ we have

\begin{enumerate}
\item[(i)] From (\ref {H1-bound}):
$$
\|u_m\|^2_{L^\infty([0,T];V)} \leq \frac{k_1} {\alpha_1^2}\quad
\hbox{or} \quad  \|v_m\|^2_{L^\infty([0,T];V\sp)} \leq k_1
$$

\item[(ii)] From (\ref {H2-ave}) we have
$$
\|u_m\|^2_{L^2([0,T];D(A))} \leq \frac{  \bar k_2(T)}{\nu \alpha_1^2}
\quad \hbox{or} \quad \|v_m\|^2_{L^2([0,T],H)} \frac{\bar k_2(T)}{\nu}
\;.
$$

\item[(iii)] From (\ref  {H2-bound})
$$
\|u_m\|^2_{L^\infty([\tau,T];D(A))} \leq \frac{ \tilde
k_2(\tau)}{\alpha_1^2} \quad \hbox{or} \quad
\|v_m\|^2_{L^\infty([\tau,T];H)} \leq \tilde
k_2(\tau) \; ,
$$
for any $\tau \in (0,T]$, where  $\tilde k_2(\tau)\to\infty $ as $\tau\to 0^+$.

\end{enumerate}

 Next, we establish uniform estimates, in $m$, for
$\frac{du_m}{ dt}$ and $\frac{dv_m}{ dt}$.

Recall~(\ref {Gal})
$$
\frac{d}{ dt} v_m(t) = -P_m \tilde B (u_m,v_m) - \nu  Av_m +
P_m f\;.
$$
{F}rom the above estimates  and part (v) of Lemma~\ref{Properties}
we have
$$
\|Av_m\|^2_{L^2([0,T],D(A)\sp)} \leq \frac {c  \bar k_2(T)}{ \nu }\; ,
$$
and
$$
\|P_m \tilde B(u_m,v_m)\|_{D(A)\sp} \leq c|u_m|^{1/2}
\|u_m\|^{1/2} |v_m| + \frac{c }{ \lambda ^{1/4}_1} |v_m| \|u_m\|\;.
$$
Consequently
$$
\|P_m \tilde B(u_m,v_m)\|^2_{L^2([0,T],D(A)\sp)} \leq
\frac {c k_1 {\bar k}_2(T)}{\nu\lambda_1^{1/2}\alpha_1^2 }
 \;.
$$
Therefore
$$
 \|\frac{dv_m }{dt}\|^2_{L^2([0,T];D(A)\sp)} \leq \tilde
k(T) \; ,
$$
and in particular
$$
\|\frac{du_m}{ dt}\|^2_{L^2([0,T],H)} \leq \frac {\tilde k(T)}{ \alpha_1^{2}}
\; ,
$$
where ${\tilde k(T)}$ is a constant which depends on $\nu,
\lambda_1,f,\alpha_0,
\alpha_1$ and $T$.

By Aubin's Compactness Theorem (see, e.g., Constantin and Foias [1988]
and Lions [1969]) we conclude that there is a
subsequence $u_{m\sp}(t)$ such that

\bigskip

$
u_{m\sp}\to u(t)  \quad {\hbox{ weakly in}}  \quad  L^2([0,T],D(A))\;,
$

$
u_{m\sp}\to u(t) \quad   {\hbox {strongly in}} \quad   L^2([0,T],V),
\quad {\hbox {and}}
$

$
u_{m\sp}\to u \quad  {\hbox {in}} \quad  C([0,T],H)\; ;
$

\bigskip

\noindent  or equivalently

\bigskip

$
v_{m\sp}\to v \quad  {\hbox  {weakly in}}  \quad L^2([0,T],H) \;,
$

$v_{m\sp}\to v \quad  {\hbox  {strongly in}} \quad  L^2([0,T],V\sp)\;,
\quad {\hbox  { and}}
$

$ v_{m\sp}\to v \quad  {\hbox   {in}} \quad  C([0,T],D(A)\sp)\;,
$

\smallskip
\noindent where $v$ is given in~(\ref{v}).

Let us relabel $u_{m\sp}$ and $v_{m\sp}$ by $u_m$ and $v_m$
respectively.  Let $w\in D(A)$, then from~(\ref  {Gal}) we have
\begin{equation}
\begin{align}
(v_m(t),w) + \nu \int^t_{t_0} (v_m(s),Aw)ds  + & \int^t_{t_0}  (\tilde
B  (u_m(s), v_m(s), P_mw)ds = \nonumber \\
& (v_m(t_0),w) + (f,P_mw) (t - t_0)\; , \nonumber
\end{align}
\end{equation}
for all $t_0, t \in [0,T]$.
Since $v_m\to v$ weakly in $L^2([0,T];H)$ then $v_m(s)\to v(s)$
weakly in $H$, for every $s\in [0,T]\backslash E$, where $|E| = 0$.
In particular, there is a subsequence of $v_m$, which we will also
denote $v_m$, such that $v_m(s) \to v(s)$ strongly in $V\sp$ and
$D(A)\sp$ for every $s\notin E$.

Now, it is clear that
$$\lim_{m\to\infty } \int^t_{t_0} (v_m(s),Aw)ds = \int^t_{t_0}
(v(s),Aw)ds\;,$$
also that $\lim\limits_{m\to\infty} |P_mAw - Aw| =
\lim\limits_{m\to\infty}|w-w_m| = 0$. On the
other hand
$$
|\int^t_{t_0} (\tilde B(u_m(s),v_m(s)),P_mw) - \langle\tilde
B(u(s),v(s),w(s)\rangle_{D(A)\sp} ds|
\leq I_m^{(1)} + I_m^{(2)} + I_m^{(3)} \; .
$$
$$
I_m^{(1)} = |\int^t_{t_0} \langle
\tilde B(u_m(s),v_m(s)),P_mw(s) - w(s)\rangle_{D(A)\sp}ds|
$$
by part (v) of  Lemma~\ref {Properties}  we have
$$
I_m^{(1)} \leq \frac {c}{ \lambda_1^{1/4}}  \int^t_{t_0} (\|u_m(s)\|
|v_m(s)|~ |P_mAw - Aw|)ds  \;,
$$
applying  Cauchy--Schwarz inequality
$$
I_m^{(1)} \leq \frac{c}{\lambda_1^{1/4}}
\left(\int^T_0 \|u_m(s)\|^2ds \right)^{1/2}
\left (\int^T_0 \|v_m(s)\|^2ds\right)^{1/2} |P_mAw - Aw|\; ,$$
and hence $\lim\limits_{m\to\infty } I_m^{(1)} = 0$.
$$
I_m^{(2)} = |\int^t_{t_0}
\langle\tilde B(u_m(s) - u_m(s), v_m(s), w\rangle_{D(A)\sp}ds| \; .
$$
Again thanks  to part (v) of Lemma~\ref  {Properties}
$$
I_m^{(2)} \leq \frac{ c}{\lambda _1^{1/4} }\int^t_{t_0} \|u_m(s) - u(s)\|
~|v_m(s)| ~|Aw|ds
$$
and by Cauchy-Schwarz
$$I_m^{(2)} \leq  \frac{ c}{\lambda _1^{1/4} } \left (\int^T_0 \|u_m(s) -
u(s)\|^2 ds\right)^{1/2} \left(\int^T_0 |v_m(s)|^2ds\right)^{1/2}
|Aw| \; .$$
Since $v_m$ bounded in $L^2([0,T];H)$ and $u_m\to u$ in
$L^2([0,T],V)$ we conclude that
$$\lim_{m\to\infty } I_m^{(2)} = 0\;.$$

Finally,
$$I_m^{(3)} = |\int^t_{t_0} \langle  \tilde B(u,v -
v_m),w\rangle_{V\sp}ds|\;,$$
by virtue of part (v) in Lemma~\ref {Properties}, and since
$v_m\to v$ weakly in $L^2([0,T];H)$, we obtain
$$\lim_{m\to\infty } I_m^{(3)} = 0\;.$$
As a result of the above we have for every $t_0,t\in [0,T]\backslash E$
\begin{equation}
\label{weak3}
\begin{align}
(v (t),w)    + \nu  \int^t_{t_0} (v(s),Aw)ds
&+\int^t_{t_0}  \langle\tilde B (u(s),v(s),w\rangle_{D(A)\sp}ds = \nonumber \\
  &(v(t_0),w) + (f,w)(t-t_0) \; ,
\end{align}
\end{equation}
for every $w \in D(A)$.
Notice that since $\|v_m(t)\|_{L^\infty ([0,T],V\sp)} \leq k_1$,
and since $v_m(t) \to v(t)$ strongly in $V\sp$ for every $t \in [0,T]
\backslash E$, we have $\|v(t)\|_{L^\infty ([0,T],V\sp)} \leq k_1$.
Moreover, because  $D(A)$ is dense in $V\sp$,~(\ref{weak3})
implies that $v(t)\in C([0,T];V\sp)$ or equivalently $u(t)\in C([0,T],V)$.

In particular, from~(\ref{weak3})  we conclude the existence of
a regular solution for the system~(\ref{The-eqn1}).

\bigskip

\noindent{\it  \underbar{Uniqueness of regular solutions}}

Next we will show the continuous dependence of regular solutions on
the initial data and, in particular, we show the uniqueness of regular
solutions.

Let $u$ and $\bar u$ be any two solutions of equation~(\ref{The-eqn1})
on the interval $[0,T]$, with initial values $u(0) = u^{in}$ and $\bar u(0) =
\bar u^{in}$ respectively.  Let us denote
$v = (\alpha_0^2 u + \alpha_1^2 Au)$, $\bar v = (\alpha_0^2\bar u +
\alpha_1^2  A\bar u)$,  $\delta u = u - \bar u$,  and by
$\delta v = v - \bar v$.  Then from
equation~(\ref{The-eqn1}) we get:
$$
\frac{d}{ dt} v + \nu  Av + \tilde B(\delta  u,v) + \tilde B(\bar
u,\delta  v) = 0\;.
$$
The above equation holds in $L^2([0,T],D(A)\sp)$, since $\delta
u$ belongs to $ L^2 ([0,T],D(A))$, the dual space of $L^2([0,T],D(A)\sp)$,
we use  Lemma~\ref {Properties} to obtain
$$
\langle\frac{d}{ dt}v,\delta u\rangle_{D(A)\sp} + \nu
(\alpha_0^2\|\delta u\|^2 + \alpha_1^2|A\delta  u|^2) + \langle \tilde
B (\bar u,\delta v), \delta u\rangle_{D(A)\sp} = 0\;.
$$
Notice that $\left\langle \frac{dv}{ dt},\delta
u\right\rangle_{D(A)\sp} = \frac{1}{ 2} \frac{d}{ dt} (\alpha_0^2 |\delta
u|^2 + \alpha_1^2\|\delta u\|^2)$, (see, e.g.,
Temam [1984],  Chapter III, Lemma 1.2).
As a result we have:
\begin{equation}
\begin{align}
\frac{1}{ 2} \frac{d}{ dt} (\alpha_0^2|\delta u|^2 + \alpha_1^2\|\delta
u\|^2) + \nu (& \alpha_0^2\|\delta u\|^2 + \alpha_1^2|A\delta u|^2) +
\nonumber \\
&+ \langle\tilde B(\bar u,\delta v),\delta u\rangle_{D(A)\sp} =0 \,.
\nonumber
\end{align}
\end{equation}
Now we use part (vi) of Lemma~\ref  {Properties} to get
\begin{equation}
\begin{align}
\frac{1}{ 2} \frac{d}{ dt} (\alpha_0^2|\delta u|^2 &+ \alpha_1^2\|\delta
u\|^2) + \nu ( \alpha_0^2\|\delta u\|^2 + \alpha_1^2|A\delta u|^2) \leq
\nonumber \\
& c(\|\bar u\|^{1/2} |A\bar u|^{1/2} |\delta v| ~\|\delta u| +
|A\bar u| ~|\delta v| ~|\delta u|^{1/2}\|\delta u\|^{1/2}) \nonumber
\end{align}
\end{equation}
and by Young's inequality we have:
\begin{equation}
\begin{align}
\frac{1}{ 2} \frac{d}{ dt} (\alpha_0^2|\delta u|^2 &+ \alpha_1^2\|\delta
u\|^2) + \nu ( \alpha_0^2\|\delta u\|^2 + \alpha_1^2|A\delta u|^2) \leq
\nonumber \\
&\leq \frac{c}{\nu } (\|\bar u\| ~|A\bar u| ~\|\delta u\|^2
+ |A\bar u|^2 |\delta u| ~\|\delta u\|) + \nonumber \\
&\quad + \frac{\nu}{ 2} (\alpha_0^2 \|\delta u\|^2 + \alpha_1^2 |A\delta
u|^2) \nonumber \\
&\leq \frac {c}{ 2\nu  \alpha_1^2 \lambda_1^{1/2}}
|A\bar u|^2 (\alpha_0^2 |\delta u|^2 + \alpha_1^2 \|\delta
u\|^2) + \nonumber \\
&\quad+ \frac{\nu }{ 2} (\alpha_0^2 \|\delta  u\|^2 + \alpha_1^2
|A\delta u|^2)\;. \nonumber
\end{align}
\end{equation}
Hence,
$$
(\alpha_0^2 |\delta u(t)|^2 + \alpha_1^2\|\delta u(t)\|^2) \leq
(\alpha_0^2 |\delta u(0)|^2 + \alpha_1^2\|\delta u(0)\|^2)
\exp\left({\displaystyle \int^t_0 \frac {c|A\bar u(s)|^2}{ \nu  \alpha_1^2
\lambda_1^{1/2}}
ds} \right ) \;.
$$
Since $\bar u\in L^2([0,T],D(A))$ we conclude the continuous dependence
of the solutions of~(\ref{The-eqn1})
on the initial data on any bounded interval $[0,T]$.
In particular, we conclude the uniqueness of regular solutions.

\end{proof}

\begin{remark}
Following the techniques introduced in \cite{Foias-Temam-Gevrey}
(see also \cite{Ferrari-Titi} and \cite{Kreiss}) we can easily show that if
the forcing  term, $f$, in equations~(\ref{The-eqn1}) belongs to a certain
Gevrey class of  regularity then the solutions of~(\ref{The-eqn1})  will
instantaneously belong to a similar Gevrey class of regularity. Specifically,
in this situation the  solution will become analytic in space and time. In
particular, one can also  provide uniform lower bounds for the radii of
analyticity (in space and  in time) for the solutions that lie in the global
attractor (see section~\ref{attractor} for the existence of a compact
finite dimensional global attractor.) As a
result of this Gevrey regularity one can also show that the Galerkin
approximating solutions, introduced earlier, converge exponentially fast in
the wave number $m$, as $m \to \infty$ (see, e.g., \cite{Doelman-Titi},
\cite{Graham-Steen-Titi},  and \cite{Jones-Margolin-Titi}). Furthermore, one
can use this Gevrey result to establish rigorous estimates for the dissipative
small scales in  equation~(\ref{The-eqn1}) (see, e.g., \cite{Doering-Titi}).

\end{remark}

\bigskip

\section{Estimating the dimension of the global attractor}
\label{attractor}

Let $S(t)$ denote the semi-group of the solution operator to
equation~(\ref{The-eqn1}), i.e. $u(t) =S(t) u^{in}$. It can
be easily shown, from the
proof of Theorem~\ref {Global}
and Rellich's Lemma (see \cite{Adams}), that   $S(t)$ is a compact
semi-group. Let us recall (see~(\ref{R})) that the ball
$B_1= \{ u \in V: \quad \|u\| \leq \frac{R_0}{\alpha_1}\}$ is an
absorbing ball, in the space $V$. Consequently, the
equation~(\ref{The-eqn1}) has a nonempty compact global attractor
\[
\cA = \cap_{s>0}\left ( \cup_{t \geq s}S(t) B_1 \right ) \;
\]
(see, e.g., \cite{Constantin-Foias-book}, \cite{Hale} and
\cite{Temam-book3}).

In this section we employ the trace formula (see, e.g.,
\cite{Constantin-Foias-85}, \cite{Constantin-Foias-book}, and
\cite{Temam-book3}) to
 estimate the Hausdorff and fractal
(box counting) dimensions of  the global attractor $\cA$ in terms of
the physical parameters of the equation~(\ref{CHeqn}). First, let us
recall the Lieb--Thirring inequality

\medskip

\begin{lemma}[The Lieb-Thirring inequality]
Let $\{\psi_j\}^N_{j=1}$ be an orthonormal set of functions in
$(H)^k
= \underbrace{H \oplus\cdots\oplus H}_{k-{\hbox{times}}}$.
Then there exists a constant
$C_{LT}$, which depends on $k$, but independent of $N$ such that
\begin{equation}
\label{Lieb-Thirring}
\int_\Omega (\sum^N_{j=1} \psi_j(x)\cdot
\psi_j(x))^{5/3} dx \leq C_{LT} \sum^N_{j=1}\int_\Omega (\nabla
\psi_j(x) : \nabla \psi_j(x))dx \;.
\end{equation}
\end{lemma}

\medskip

Next we will present a new technical Lemma which we will use in estimating
the dimension of the global attractor.

\medskip

\begin{lemma}   Let $\{\varphi_j\}^N_{j=1}\subset V$ be an
orthonormal set with respect to the inner product $[\cdot,\cdot]$ which is
defined in~(\ref{inner}), i.e.
$$[\varphi_i,\varphi_j] = \alpha_0^2(\varphi_i,\varphi_j) +
\alpha_1^2((\varphi_i,\varphi_j)) = \delta_{ij}\;.
$$
Let $\psi_j(x) = (\alpha_0\varphi_j(x), \alpha_1
\frac{\partial}{\partial x_1} \varphi_j(x), \alpha_1
\frac{\partial}{\partial x_2} \varphi_j(x), \alpha_1
\frac{\partial}{\partial x_3} \varphi_j(x))^T\;,$ and \\
 $\varphi^2(x) = \sum^N_{j=1} (\varphi_j(x)\cdot \varphi_j(x))$.
Then, there exists a constant $C_F$, which is independent of $N$, such
that
\begin{equation}
\label{L-Infinity}
\|\varphi(x)\|^2_{L^\infty } \leq \frac{C_F}{ \alpha_1^2}
\left(\sum^N_{j=1} \int_\Omega (\nabla \psi_j(x): \nabla
\psi_j(x))dx \right)^{1/2}\;.
\end{equation}
\end{lemma}
\begin{proof}
 Let $\xi_j\in\bR\;,\;j=1,\ldots,N$, to be
chosen later. By Agmon's inequality~(\ref{Agmon}) we have
\begin{equation}
\begin{align}
\alpha_0^2 | \sum^N_{j=1} \xi_j (A^{-1/2} \varphi_j)(x)|^2
+ & \alpha_1^2 | \sum^N_{j=1} \xi_j \varphi_j(x)|^2 \leq \nonumber \\
& c\alpha_0^2 |\sum^N_{j=1} \xi_j \varphi_j|
~\| \sum^N_{j=1} \xi_j \varphi_j\| +
c\alpha_1^2 \| \sum^N_{j=1} \xi_j \varphi_j\|
~|\sum^N_{j=1} \xi_j A\varphi_j| \;, \nonumber
\end{align}
\end{equation}
and by Cauchy-Schwarz
\begin{equation}
\begin{align}
&\alpha_0^2 | \sum^N_{j=1} \xi_j (A^{-1/2} \varphi_j)(x)|^2
+  \alpha_1^2 | \sum^N_{j=1} \xi_j \varphi_j(x)|^2  \nonumber \\
&\leq c \left ( \alpha_0^2 |\sum^N_{j=1} \xi_j \varphi_j|^2 +
\alpha_1^2 \|\sum^N_{j=1} \xi_j \varphi_j\|^2\right)^{1/2}
\left(\alpha_0^2 \|\sum^N_{j=1} \xi_j \varphi_j\|^2 +
\alpha_1^2 |\sum^N_{j=1} \xi_j A \varphi_j|^2 \right)^{1/2} \nonumber \\
&\leq c\left[ \sum^N_{j=1} \xi_j \varphi_j,
\sum^N_{j=1} \xi_j \varphi_j \right]^{1/2}
\left( \sum^N_{j=1} \xi^2_j \right)^{1/2}
\left(\alpha_0^2 \sum^N_{j=1} \|\xi_j\|^2 +
\alpha_1^2 \sum^N_{j=1} |A\varphi_j|^2\right)^{1/2} \;. \nonumber
\end{align}
\end{equation}
Since $[\varphi_i,\varphi_j] = \delta _{ij}$ we have
\begin{equation}
\begin{align}
\alpha_0^2 | \sum^N_{j=1} \xi_j (A^{-1/2} \varphi_j)(x)|^2
+  &\alpha_1^2 | \sum^N_{j=1} \xi_j \varphi_j(x)|^2  \nonumber \\
&\leq c(\sum^N_{j=1} \xi^2_j)
\left (\alpha_0^2 \sum^N_{j=1} \|\varphi_j\|^2 +
\alpha_1^2 \sum^N_{j=1} |A \varphi_j|^2\right)^{1/2}
\nonumber \\
&\leq c(\sum^N_{j=1} \xi^2_j)
\left (\sum^N_{j=1}\int_\Omega (\nabla \psi_j :
\nabla \psi_j)dx \right)^{1/2} \;.
\nonumber
\end{align}
\end{equation}
Let $i \in \{1,2,3\}$ be fixed, we choose $\xi_j = \varphi_{ji}(x)$,
from the above we have
$$
\alpha_1^2(\sum^N_{j=1} \varphi^2_{ji}(x))^2
\leq c(\sum^N_{j=1} \varphi^2_{ji}(x))
\left (\sum^N_{j=1}\int_\Omega (\nabla \psi_j(x) : \nabla\psi_j (x))dx
\right )^{1/2} \;.
$$
Now we sum over $i$, $i=1,2,3$,  to reach
$$
\alpha_1^2 \varphi^2(x) \leq \left (\sum^N_{j=1} \int_\Omega (\nabla
\psi_j(x) : \nabla \psi_j (x))dx\right )^{1/2} \;,
$$
which concludes our proof.

\end{proof}

\begin{theorem}
\label{dimension-Grashoff}   The Hausdorff and fractal dimensions
of the global attractor of the viscous
Camassa--Holm (NS$-\alpha$) equations, $d_H(\cA)$ and $d_F(\cA)$, respectively,
satisfy:
$$  d_H(\cA) \le d_F(\cA) \leq c\max \left \{G^{4/3}
(\frac{1}{ \gamma \alpha_1^2\lambda_1})^{2/3} , G^{3/2}
(\frac{1}{ \alpha_0^4 \gamma^2\lambda_1\alpha_1^2})^{3/8} \right \} \;,
$$
where $G = \frac{|f|}{ \nu ^2\lambda_1^{3/4}}$ is the Grashoff number
and, as before,   $\frac{1}{\gamma } = \min\left\{ \frac{1}{\alpha_0^2},
\frac{1}{ \alpha_1^2\lambda_1}\right\}$.

\end{theorem}

\begin{proof}  The linearized~(\ref{The-eqn1}) equation about a
regular solution $u(t)$ takes
the form
\begin{equation}
\label{linear1}
\frac{d}{ dt}\delta v + \nu A\delta v + \tilde B(\delta u,v) + \tilde
B (u,\delta v) = 0\;,
\end{equation}
where $ v (t) = \alpha_0^2   u + \alpha_1^2 A  u$ and
$\delta v = \alpha_0^2 \delta u + \alpha_1^2 A\delta u$.  Notice
that $\delta u$ evolves according to the  equation
\begin{equation}
\label{linear2}
\frac{d}{dt} \delta u + \nu  A\delta u + (\alpha_0^2 I +
 \alpha^2_1 A)^{-1} [\tilde B(\delta u, \alpha^2_0u + \alpha^2_1Au) + \tilde
B(u,\alpha^2_0\delta u + \alpha^2_1A\delta u) = 0\;,
\end{equation}
which we write symbolically  as
$$
\frac{d}{ dt} \delta u + T(t)\delta u = 0 \;.
$$

Let $\delta u_j(0)$, for $j=1,\ldots,N$,  be a set of linearly
independent vectors in $V$, and let $\delta u_j(t)$
be the corresponding solutions of~(\ref{linear2}) with initial value
$\delta u_j(0)$, for $j=1,\ldots,N$.  We denote
\begin{equation}
\label{trace1}
\cT_N (t) = \;{\rm Trace}\; ( P_N(t)T(t)\big|_{P_N(t)V}) \;,
\end{equation}
where $P_N(t)V = \bR\delta v_1(t) + \bR\delta v_2(t)
+\cdots+ \bR\delta v_N(t)$,
and $P_N(t)$ is the orthogonal projector of $V$ onto $P_N(t)V$ with
respect to the inner product $[\cdot,\cdot]$ given in~(\ref{inner}).

Let $\{\varphi_j(t)\}^N_{j=1}$ be an orthonormal basis, with respect to
inner product $[\cdot,\cdot]$, of the space $P_NV$, i.e.
$[\varphi_i,\varphi_j] = \delta _{ij}\;,\;i,j=1,\ldots,N$. We set
$$
\psi_j = (\alpha _0 \varphi_j, \alpha _1
\frac{\partial}{\partial x_1} \varphi_j, \alpha _1
 \frac{\partial}{\partial x_2}\varphi_j, \alpha _1
 \frac{\partial}{\partial x_3}\varphi_j)^T\;.
$$
Notice that $(\psi_j,\psi_k) = \delta _{jk}\;,\; j,k=1,\ldots,N$.
We set
\begin{equation}
\begin{align}
\psi^2(x,t) &= \sum^N_{j=1} (\psi_j(x,t)\cdot \psi_j(x,t) )  \nonumber \\
&= \alpha_0^2 \sum^N_{j=1} \varphi_j(x,t) \cdot \varphi_j(x,t) +
\alpha^2_1 \sum^N_{j=1} ( \nabla \varphi_j (x,t) : \nabla
\varphi_j(x,t) ) \nonumber
\end{align}
\end{equation}
Notice that by the Lieb-Thirring inequality~(\ref{Lieb-Thirring})
$$
\int_\Omega (\psi(x,t))^{10/3}dx \leq C_{LT} Q_N(t)\;,
$$
where $Q_N(t) := \sum^N_{j=1} \int_\Omega (\nabla \psi_j(x,t) :
\nabla\psi_j(x,t))dx$.

Let us denote $\theta_j(x,t) = \alpha^2_0 \varphi_j(x,t) +
\alpha^2_1A\varphi_j(x,t)$, for $j = 1,\ldots,N$. From~(\ref{trace1})
we have
\begin{equation}
\begin{align}
\cT_N(t) &= \sum^N_{j=1}
[T(t)\varphi_j(\cdot,t),\varphi_j(\cdot,t)]  \nonumber \\
&= \nu  \sum^N_{j=1} [A\varphi_j,\varphi_j] + \sum^N_{j=1}
(\tilde B(\varphi_j,v), \varphi_j) + \sum^N_{j=1}
(\tilde B(u,\theta_j),\varphi_j) \;, \nonumber
\end{align}
\end{equation}
and by virtue of~(\ref{BI2}) we have
\[
\cT_N(t)
= \nu  \sum^N_{j=1} [A\varphi_j,\varphi_j] + \sum^N_{j=1}
(\tilde B(u,\theta_j), \varphi_j)\;.
\]
Observe that
\begin{equation}
\begin{align}
\sum^N_{j=1} [A\varphi_j,\varphi_j] &= \alpha^2_0 \sum^N_{j=1}
(A\varphi_j,\varphi_j) + \alpha^2_1 \sum^N_{j=1}
(A\varphi_j,A\varphi_j) =  \nonumber \\
&= \sum^N_{j=1} \int_\Omega (\nabla \psi_j (x,t): \nabla \psi_j(x,t))dx =
Q_N(t) \;. \nonumber
\end{align}
\end{equation}
Thus
\begin{equation}
\label{trace2}
\cT_N(t)
= \nu Q_N(t) + \cJ_N(t)\; ,
\end{equation}
where $\cJ_N(t) := \sum^N_{j=1} (\tilde
B(u,\theta_j),\varphi_j)$. Let us now estimate $\cJ_N(t)$. Using~(\ref{BI2})
and~(\ref{BI1}) we have
\begin{equation}
\begin{align}
\cJ_N(t) &= \sum^N_{j=1} [((u\cdot\nabla)\theta_j,\varphi_j) +
((\varphi_j\cdot\nabla)u,\theta_j)] \nonumber \\
&= \sum^N_{j=1} [-((u\cdot\nabla)\varphi_j,\theta_j) +
((\varphi_j\cdot\nabla)u,\theta_j)] \;, \nonumber
\end{align}
\end{equation}
again by~(\ref{BI1})
\[
\cJ_N(t) = \sum^N_{j=1} [-\alpha^2_1((u\cdot\nabla)\varphi_j,A\varphi_j) +
\alpha^2_0((\varphi_j\cdot\nabla)u,\varphi_j) + \alpha_1^2
((\varphi_j\cdot\nabla)u,A\varphi_j)   \;,
\]
integrating by  parts and using~(\ref{BI1})
\begin{equation}
\begin{align}
\cJ_N(t)
&= \sum^N_{j=1} \sum^3_{k=1} \alpha_1^2
((\frac{\partial}{\partial x_k}u\cdot \nabla)
\varphi_j,\frac{\partial}{\partial x_k}\varphi_j) +
\alpha^2_0 \sum^N_{j=1} ((\varphi_j\cdot \nabla)u,\varphi_j) - \nonumber \\
&- \alpha^2_1 \sum^N_{j=1} \sum^3_{k=1}
((\frac{\partial}{\partial x_k}u\cdot \nabla)u,
\frac{\partial}{\partial x_k} \varphi_j) -
\alpha^2_1 \sum^N_{j=1} \sum^3_{k=1}
((\varphi_j\cdot \nabla)
\frac{\partial}{\partial x_k}u, \frac{\partial}{\partial x_k}\varphi_j) \;.
 \nonumber
\end{align}
\end{equation}
Therefore,
\begin{equation}
\begin{align}
& |\cJ_N(t)| \leq c\int_\Omega (\nabla u(x,t) : \nabla u(x,t))^{1/2}
\psi^2(x,t)dx +  \nonumber \\
&+ c\alpha^2_1 \int_\Omega \left [(\sum^3_{i,k=1}
(\frac{\partial^2}{\partial x_i\partial
x_k}u(x,t))^2) \varphi^2(x,t) (\sum^N_{j=1}
 (\nabla \varphi_j(x,t) : \nabla \varphi_j(x,t)))\right ]^{1/2}dx
\;,
 \nonumber
\end{align}
\end{equation}
where $\varphi^2(x,t) = \sum^N_{j=1} (\varphi_j(x,t)\cdot
\varphi_j(x,t))$. As a result we have
\begin{equation}
\begin{align}
|\cJ_N(t)| &\leq c \int_\Omega (\nabla u(x,t): \nabla u(x,t))^{1/2}
\psi^2 (x,t)dx +\nonumber \\
&+ {\alpha_1} \int_\Omega \psi(x,t)\varphi(x,t)
(\sum^3_{k,i=1} (\frac{\partial^2}{\partial x_i\partial x_k}
u(x,t))^2)^{1/2} dx\;.
\nonumber
\end{align}
\end{equation}
Thanks to~(\ref{L-Infinity}) we have
\begin{equation}
\label{J}
\begin{align}
|\cJ_N(t)| \leq c \int_\Omega (\nabla u(x,t): \nabla u(x,t))^{1/2}
\psi^2 (x,t)dx + &  \nonumber  \\
+  C_F^{1/2} Q_N^{1/4}(t) \int_\Omega
\left(\sum^3_{k,i=1} (\frac{\partial^2}{\partial x_i\partial x_k}
u(x,t))^2 \right)^{1/2} \psi(x,t)dx\; , &
\end{align}
\end{equation}
and by the H\"{o}lder inequality we get
\[
|\cJ_N(t)| \leq c \|\nabla u\|_{L^{5/2}}
\left(\int_\Omega (\psi(x,t))^{10/3}dx\right)^{3/5} +
c Q_N^{1/4}(t) |Au| \left( \int_\Omega \psi^2(x,t)dx \right)^{1/2} \;.
\]
Since $[\varphi_i,\varphi_j]= \delta_{ij}$ we have
$\int_\Omega\psi^2(x,t)dx =N$. Therefore, the above gives
\[ |\cJ_N(t)| \leq c\|\nabla u\|_{L^{5/2}}
\left(\int_\Omega(\psi(x,t))^{10/3}dx \right)^{3/5} + cQ_N^{1/4}(t)
|Au| N^{1/2}\;.
\]
Applying the Lieb--Thirring inequality~(\ref{Lieb-Thirring}) we obtain
\begin{equation}
\begin{align}
|\cJ_N(t)| \leq c~c_{LT} & \|\nabla u\|_{L^{5/2}}
\left (\sum^N_{j=1} \int_\Omega (\nabla \psi_j(x,t):
\nabla\psi_j(x,t))dx\right )^{3/5} + \nonumber \\
&\quad+ c Q_N^{1/4}(t) |Au| N^{1/2} \;,
\nonumber
\end{align}
\end{equation}
that is
\[
|\cJ_N(t)| \leq c\|\nabla u\|_{L^{5/2}}
Q_N^{3/5}(t) + cQ_N^{1/4}(t) |Au| N^{1/2} \;.
\]
Applying  Young's inequality
$$
|\cJ_N(t)| \leq \frac{c}{\nu ^{3/2}} \|\nabla u\|_{L^{5/2}}^{5/2} +
\frac{\nu }{ 2} Q_N(t) + c|Au|^{4/3}
\left(\frac{N^2}{ \nu }\right)^{1/3}\;,
$$
then using H\"older's inequality
$$
|\cJ_N(t)| \leq \frac {c}{ \nu^{3/2}} \|\nabla u\|_{L^2}^{7/4}
\|\nabla u\|^{3/4}_{L^6} + \frac{\nu }{ 2} Q_N(t) + c|Au|^{4/3}
\left(\frac{N^2}{ \nu }\right)^{1/3}\;,
$$
and by virtue of the Sobolev inequality~(\ref{Sobolev}c) we obtain
$$
|\cJ_N(t)| \leq \frac {c}{ \nu^{3/2}} \| u\|^{1/2} \| u\|^{5/4}|Au|^{3/4}
+ \frac{\nu }{ 2} Q_N(t) + c|Au|^{4/3}
\left(\frac{N^2}{ \nu }\right)^{1/3}\;.
$$
Using Young's inequality again we reach
$$
|\cJ_N(t)| \leq \frac{c}{ \nu^{3/2}} \frac {\|u\|^{1/2} }{
\alpha_0^{5/4}\alpha _1^{3/4}} (\alpha_0^2 \|u\|^2 + \alpha^2_1 |Au|^2) +
\frac{\nu }{ 2} Q_N(t) + c|Au|^{4/3}
\left(\frac{N^2}{ \nu }\right)^{1/3} \;.
$$
Substituting in~(\ref{trace2}) we get:
\begin{equation}
\label{trace3}
\begin{align}
\cT_N(t) \geq &\frac{\nu}{ 2} Q_N(t) - \frac {c}{ \nu ^{3/2}}
\frac{\|u\|^{1/2}}{ \alpha_0^{5/4} \alpha_1^{3/4}} (\alpha^2_0\|u\|^2
+ \alpha^2_1|Au|^2) - \nonumber \\
&- c|Au|^{4/3} \left(\frac{N^2}{ \nu }\right)^{1/3} \;.
\nonumber
\end{align}
\end{equation}
Now, we require  $N$ to be large enough such that
\begin{equation}
\label{Mean-trace}
\liminf_{T\to\infty } \frac{1}{ T} \int^{T}_0 \cT_N(s)ds > 0\;.
\end{equation}
According to the trace formula (see \cite{Constantin-Foias-85},
 \cite{Constantin-Foias-book} or \cite{Temam-book3}) such an $N$
will be an upper bound for the fractal and Hausdorff dimensions
of the global attractor.
Observe that from the asymptotic behavior of the eigenvalues of the
operator $A$  there is a constant $c_0$ such that
$$
\lambda _j \geq c_0\lambda_1 ~j^{2/3} \quad {\hbox{for}} \quad
 j = 1,2,\ldots \;.
$$
Therefore, since $Q_N(t)$ is the trace of the operator $A$ restricted
to some subspace of dimension $N$, we have
\begin{equation}
\label{Q}
Q_N(t) \geq \sum^N_{j=1} \lambda _j \geq c\lambda _1 N^{5/3}\;.
\end{equation}
Let us require $N$ to be large enough so that
$$
\nu \lambda _1 N^{5/3} \geq c \limsup_{T\to\infty }\left  (\frac{1}{
T} \int^{T}_0 |Au(s)|^{4/3}ds\right) \left(\frac{N^2}{ \nu }\right)^{1/3}
 $$
and
$$
\nu \lambda_1N^{5/3} \geq \frac{c}{ \nu
^{3/2} \alpha _0^{5/4} \alpha _1^{3/4}}\limsup_{T\to\infty }
\frac{1}{ T} \int^{T}_0
\|u(s)\|^{1/2} (\alpha^2_0\|u(s)\|^2 + \alpha^2_1|Au(s)|^2)ds \;.
$$
For such an $N$ the inequality~(\ref{Mean-trace}) is satisfied,
and thus  $N$ provides  an upper bound for the fractal and Hausdorff
 dimensions of the global attractor.

By H\"older's inequality we have
$$\limsup_{T\to\infty } \frac{1}{ T} \int^{T}_0 |Au(s)|^{4/3}ds \leq
\limsup_{T\to\infty } \left(\frac{1}{ T} \int^{T}_0
|Au(s)|^2\right)^{2/3}$$
and thanks to~(\ref  {Mean-R}) we get
$$
\limsup_{T\to\infty } \frac{1}{T} \int^{T}_0 |Au(s)|^{4/3}ds \leq
\left(\frac{G^2\nu ^2\lambda_1^{1/2} }{ \gamma \alpha_1^2}\right)^{2/3}
\;.
$$
Therefore, from the above, ~(\ref{R}) and~(\ref {Mean-R}) we have
$$
d_H(\cA) \leq d_F(\cA) \leq c\max\left \{ G^{4/3}
\left (\frac{1}{\gamma \alpha^2_1\lambda _1}\right)^{2/3}\;,\; G^{3/2}
\left(\frac{1}{ \alpha_0^2
\gamma ^2\lambda _1\alpha^2_1}\right)^{3/8}\right\} \;,$$
which concludes our proof.

\end{proof}

\bigskip

\section{Numbers of degrees of freedom in turbulent flows}
\label{degrees of freedom}

An argument from the classical theory of turbulence
\cite{Landau-Lifshitz} suggests that
there are finitely many degrees of freedom in turbulent
flows. Heuristic physical arguments are used to justify
this assertion and to provide an estimate for  this number
of degrees of freedom by dividing a typical length scale
of the flow, $L$, by  the Kolmogorov
dissipation length scale and taking the third power in $3D$. The
resulting formula is usually expressed explicitly in terms of the mean
rate of dissipation of energy and the kinematic viscosity.
In analogy with this heuristic approach we  will derive here
an estimate for the ``dissipation'' length scale (i.e., what
would correspond to the Kolmogorov length scale) for the viscous
Camassa--Holm  (NS$-\alpha$) equations in terms of the mean rate of
dissipation of ``energy'' and the kinematic viscosity.
We will also show that the corresponding number of
degrees of freedom is proportional to the dimension of
the global attractor.  This, in a sense, suggests that
in the absence of boundary effects (e.g., in the case
of periodic boundary conditions) the viscous 
Camassa--Holm equations represent, very well, the averaged
equation of motion of turbulent flows. Hence, one is tempted
to use the viscous Camassa--Holm equations  as a closure model for the
Reynolds equations, which represent the ensemble-averaged Navier--Stokes
equations. Indeed,  this idea motivated
  our studies in  \cite{Chen1}, \cite{Chen2} and \cite{Chen3},
and it also led to a physical derivation in \cite{Holm2}
(see also \cite{Chen2})
of the   viscous Camassa--Holm  (NS$-\alpha$) equations, in the inviscid
case, as  averaged  equations.

As before, we denote $v = \alpha_0^2u + \alpha_1^2 Au$, hence 
equation~(\ref{The-eqn1}) and equation~(\ref{The-eqn2}) take the form
\begin{eqnarray}
\label{v-eqn}
\frac {dv}{ dt} + \nu  Av + \tilde B(u,v) = f &\\
u(0) = u^{in}& \nonumber
\end{eqnarray}
We define 
\begin{equation}
\label{epsilon}
\epsilon(u^{in}) = \lambda^{3/2}_1 \nu \left [
 \limsup_{T\to\infty }  \frac{1}{ T}
\int^T_0 (\alpha^2_0 \|u(s)\|^2 + \alpha^2_1|Au(s)|^2)ds \right ] \; ,
\end{equation}
the mean rate of dissipation of ``energy'',
and 
\[
\bar{\epsilon} = \sup_{u^{in} \in \cA} \epsilon(u^{in}) \, ,
\]
the maximal mean rate of
dissipation of energy on the attractor.  From equation~(\ref{J}),
and since $\int\psi^2 (x,t)dx = N$, we have:
\begin{eqnarray*}
|\cJ_N(t)| &\leq c\int _\Omega (\nabla u(x,t): \nabla u(x,t))^{1/2}
\psi^2 (x,t)dx + \\
&+ c C_F^{1/2} Q_N^{1/4} (t) |Au| N^{1/2}\;,
\end{eqnarray*}
and by H\"older's inequality we have
\[
|\cJ_N(t)| \leq c \|\nabla u\|_{L^6} \|\psi^2\|_{L^{6/5}} +
cQ_N^{1/4} (t) |Au| N^{1/2}.
\]
Again by H\"older's inequality and~(\ref {Sobolev}) we get
\begin{eqnarray*}
|\cJ_N(t)| &\leq c |Au(t)| \left(\int_\Omega \psi^2 (x,t)dx \right)^{7/{12}}
~ \left(\int_\Omega (\psi^2(x,t))^{5/3}dx\right)^{1/4} + \\
&+ cQ_N^{1/4}(t)
|Au(t)| N^{1/2}\;.
\end{eqnarray*}
Using the Lieb--Thirring inequality~(\ref{Lieb-Thirring}) and 
$\int \psi^2 (x,t)dx = N$
we obtain
\[
|\cJ_N(t)| \leq c |Au(t)| N^{7/12} Q_N^{1/4}(t) + cQ_N^{1/4}(t) |Au(t)|
N^{1/2} ,
\]
and, hence
\[
|\cJ_N(t)| \leq c|Au(t)| N^{7/12} Q_N^{1/4}(t)\;.
\]
After applying Young's inequality to the above and substituting
in  equation~(\ref{trace2}) we obtain
\[
\cT_N(t) \ge  \frac{\nu}{2} Q_N (t) - c \nu^{-1/3} N^{7/9} |Au(t)|^{4/3}\; .
\]
Therefore, in order to satisfy~(\ref  {Mean-trace}),  and based on the
above,  it suffices to
choose $N$ large enough so that for every trajectory $ u(t)$ on the global
attractor $\cA$ we have
\[
\liminf_{T\to\infty }  \frac{1}{ T} \int^T_0 [ \frac{\nu}{2} Q_N (t)
 - c \nu^{-1/3} N^{7/9} |Au(t)|^{4/3}]dt > 0\;.
\]
Therefore,  such a large  $N$ is  an upper bound for
the dimension of the global
attractor. Based on~(\ref{Q}) it
suffices to require
\[
\nu^{4/3} \lambda_1 N^{5/3} \cdot N^{-7/9} >
c ~\limsup_{T\to\infty } \frac{1}{ T} \int^T_0 |Au(s)|^{4/3}ds,
\]
for every solution $u^{in} \in \cA$, i.e.,
\[
\nu^{4/3} \lambda_1 N^{8/9} > c
\sup_{u^{in} \in \cA} \left ( \limsup_{T\to\infty }
\frac{1}{ T} \int^T_0 |Au(s)|^{4/3}ds \right ) \; .
\]
On the other hand, using H\"older's inequality  and~(\ref{epsilon})
we have
\begin{equation}
\begin{align}
\sup_{u^{in} \in \cA}
\left ( \limsup_{T\to\infty } \frac{1}{ T} \int^T_0 |Au(s)|^{4/3} ds \right )
\leq \sup_{u^{in} \in \cA}
\left (  \limsup_{T\to\infty }(\frac{1}{ T} \int^T_0 |Au(s)|^2ds \right)^{2/3}
& \nonumber \\
\leq \sup_{u^{in} \in \cA} \left (  \limsup_{T\to\infty }
(\frac{1}{ T\alpha^2_1} \int^T_0
(\alpha^2_0 \|u(s)\|^2 + \alpha^2_1 |Au(s)|^2)ds\right)^{2/3} & \nonumber \\
\leq \sup_{u^{in} \in \cA} \left (  \frac{\epsilon (u^{in})}
{ \nu \lambda _1^{1/2} (\alpha^2_1\lambda_1)}\right)^{2/3}\; \leq
 \left (  \frac{\bar \epsilon }
{ \nu \lambda _1^{1/2} (\alpha^2_1\lambda_1)}\right)^{2/3}& .  \nonumber
\end{align}
\end{equation}
Therefore,  every  $N$ large enough that
\begin{equation}
\label{freedom1}
N \geq c \left (\frac{ \bar\epsilon}
{\nu^3\lambda _1^2 (\alpha^2_1\lambda _1)}\right)^{3/4}
\end{equation}
  is an upper bound for  the  fractal
dimension of the global  attractor, and hence is an upper bound
for the number of degrees of freedom in turbulent flows.

We set the dissipation length scale, in analogy with the Kolmogorov
dissipation length scale in the classical theory of turbulence, to be
$$
\ell_\epsilon  = \left (\frac{\nu ^3} {\bar \epsilon }\right)^{1/4}.
$$
Then equation~(\ref  {freedom1}) leades to the following

\begin{theorem}
 The Hausdorff and fractal dimensions
of the global attractor of the viscous
Camassa--Holm (NS$-\alpha$) equations, $d_H(\cA)$ and $d_F(\cA)$, respectively,
satisfy:
$$  d_H(\cA) \le d_F(\cA) \leq  \frac{c}{(\alpha^2_1\lambda _1)^{3/4}}  \left(\frac{1}
{\ell_\epsilon \lambda _1^{1/2}} \right)^3\;.
$$
\end{theorem}

This estimate for the number of degrees of freedom is
consistent with the conventional estimate a la
Kolmogorov--Landau--Lifshitz \cite{Landau-Lifshitz}. In particular,
the number of degrees of freedom scales as the cube of the
ratio of the domain size divided by the Kolmogorov
dissipation length scale (times a factor involving the fixed
$\alpha_1$).

\bigskip

\section{Convergence to the Navier--Stokes equations}
\label{convergence}

We observed earlier  that the system~(\ref{The-eqn1})
reduces to the Navier--Stokes for $\alpha_0 =1$
 and $\alpha_1 =0$. In this section we will fix
$\alpha_0 =1$ and investigate the convergence of the
solutions of the system~(\ref{The-eqn1}) as
$\alpha_1 \to 0^+$, and relate the limit to the
Navier--Stokes equations. We will be studying further
the relation between solutions of the
system~(\ref{The-eqn1}) and the $3D$ Navier--Stokes
equation in a subsequent work.

\begin{theorem} Let $f\in H$, $u^{in} \in V$ and
$\alpha_0 =1$.  Let $u_{\alpha_1}$ and 
$v_{\alpha_1}= u_{\alpha_1} + \alpha_1^2 A u_{\alpha_1}$
denote the solution
of the initial-value problem~(\ref{The-eqn1}) (or equivalently~(\ref{v-eqn})).
Then there are subsequences $u_{\alpha_1^j}$,  $v_{\alpha_1^j}$, and a
function $u$ such that as $\alpha_1^j \to 0^+$ we have: \\ (i) $u_{\alpha_1^j}
\to u$, strongly in $L^2_{\mbox{loc}} ([0,\infty); H)$; \\
(ii) $u_{\alpha_1^j} \to u$,  weakly in
$L^2_{\mbox{loc}} ([0,\infty);V)$;\\
(iii) for every $T \in (0,\infty)$ and every $w \in H$ we have
$(u_{\alpha_1^j}(t),w)\to (u(t),w)$  uniformly on $[0,T]$; \\
(iv) $v_{\alpha_1^j} \to u$ weakly in $L^2_{\mbox{loc}}
([0,\infty); H)$ and strongly in $L^2_{\mbox{loc}}
([0,\infty); V\sp)$. \\
Furthermore, $u$ is a weak solution of the $3D$ Navier--Stokes
equations with the initial data $u(0) = u^{in}$ (for the definition
of weak solutions to the $3D$ Navier--Stokes equations see
\cite{Constantin-Foias-book} and \cite{Temam-book1}.)

\end{theorem}

\begin{proof}
Let $T>0$ be fixed. From the proof of Theorem~\ref{Global} and
by passing to the limit one can show
that the estimates~(\ref{H1-bound}) and~(\ref{H2-ave})  also hold  for the
exact solution  of the system~(\ref{The-eqn1}). That is
\[
|u_{\alpha_1}(t)|^2 + \alpha_1^2 \|u_{\alpha_1}(t)\|^2 \leq
k_1 \; ,
\]
and
\[
\nu  \int^{T}_0 (\|u_{\alpha_1}(s)\|^2 + \alpha_1^2
|Au_{\alpha_1}(s)|^2)ds \leq \bar k_2(T)\; .
\]
This implies  that   there are subsequences $\{u_{\alpha_1^j}\}$ and
$\{v_{\alpha_1^j}\} $, and corresponding functions $u$ and $v$
such that:
\[
\{u_{\alpha_1^j}\} \to u \qquad {\mbox {weakly in}} \qquad L^2 ([0,T]; V),
\]
and
\[
\{v_{\alpha_1^j}\} \to v \qquad {\mbox {weakly in}} \qquad L^2 ([0,T]; H),
\]
as $\alpha_1^j \to 0^+$.

Next we will use the above estimates and equation~(\ref{The-eqn1}) to show
that 
\begin{equation}
\label{dudtL2}
\int_0^T \left |A^{-1}\frac{du_{\alpha_1}(t)}{dt}\right|^{2} dt=
\int_0^T \left\|\frac{du_{\alpha_1}(t)}{dt}\right\|_{D(A)\sp}^{2} dt \leq
K(T)\; ,
\end{equation}
for some constant $K$ which depends on $T$, but is independent of $\alpha_1$.
Indeed, from equation~(\ref{The-eqn1}) (or equivalently~(\ref{v-eqn}))
 we have 
\[
\frac {du_{\alpha_1}}{ dt} + \nu  A u_{\alpha_1}+ 
(I + {\alpha_1}^2 A)^{-1} \tilde B(u_{\alpha_1},v_{\alpha_1}) = (I +
{\alpha_1}^2 A)^{-1} f \; . \]
Thus,
\[
\left |A^{-1}\frac{du_{\alpha_1}(t)}{dt}\right| \leq \nu |u_{\alpha_1}| +
| A^{-1}(I + {\alpha_1}^2 A)^{-1} \tilde B(u_{\alpha_1},v_{\alpha_1}) |  
+|A^{-1}f| \; .
\]
In order to prove~(\ref{dudtL2}) we only need to find the proper 
estimate for 
$$| A^{-1}(I + {\alpha_1}^2 A)^{-1} \tilde B(u_{\alpha_1},v_{\alpha_1}) |\le
| A^{-1} \tilde B(u_{\alpha_1},v_{\alpha_1}) | \; .
$$
Applying part (v) of Lemma~\ref{Properties} we obtain
\begin{eqnarray*}
| A^{-1} \tilde B(u_{\alpha_1},v_{\alpha_1}) | 
&\leq &
c \left (|u_{\alpha_1}|^{1/2} \|u_{\alpha_1}\|^{1/2} |v_{\alpha_1}| +
\lambda_1^{-1/4} |v_{\alpha_1}| \|u_{\alpha_1}\|  \right ) \\
&\le& 2c \lambda_1^{-1/4} |v_{\alpha_1}| \|u_{\alpha_1}\| \\
&\le&  2c  \lambda_1^{-1/4} \|u_{\alpha_1}\| ( |u_{\alpha_1}|
+ {\alpha_1}^2 |Au_{\alpha_1}|) \; .
\end{eqnarray*}
As a result of the above estimates we have
\begin{eqnarray*}
| A^{-1} \tilde B(u_{\alpha_1},v_{\alpha_1}) |^2
&\leq &
8 c^2 \lambda_1^{-1/2}\left (\|u_{\alpha_1}\|^2 |u_{\alpha_1}|^2
+  ({\alpha_1}^2 \|u_{\alpha_1}\|^2)({\alpha_1}^2|Au_{\alpha_1}|^2)\right) \\
&\leq &
8 c^2 \lambda_1^{-1/2} k_1 \left ( \|u_{\alpha_1}\|^2
+   {\alpha_1}^2|Au_{\alpha_1}|^2\right) \; ,
\end{eqnarray*}
and by integrating the above estimate over the interval $[0,T]$ we have
\[
\int_0^T | A^{-1} \tilde B(u_{\alpha_1}(t),v_{\alpha_1}(t)) |^2 dt \le
\frac{\bar k_2(T)}{\nu}8 c^2 \lambda_1^{-1/2} k_1 \; .
\]
{F}rom all the above we conclude~(\ref{dudtL2}).

By virtue of the above estimates and Aubin's compactness Theorem
(see, e.g., \cite{Constantin-Foias-book}, \cite{Lions}, or \cite{Temam-book1})
there exists a subsequence, which will also be labeled by $\{u_{\alpha_1^j}\}$,
that   converges to  $u$ strongly in $L^2 ([0,T]; H)$. Furthermore, since
\[
\int_0^T |A^{-1/2}(v_{\alpha_1^j}(t)- u_{\alpha_1^j}(t))|^2 dt =
 (\alpha_1^j)^2\int_0^T \|u_{\alpha_1^j}(t)\|^2 dt
\le (\alpha_1^j)^2 \frac{\bar k_2(T)}{\nu}
\; ,
\]
we have that $ v_{\alpha_1^j} \to u $ strongly in $L^2 ([0,T]; V\sp)$,
as $\alpha_1^j \to 0^+$;
and that $v(t) = u(t)$ a.e. in $[0,T]$.

As a result of these estimates one can extract  subsequences, which will
be also labeled  by $\{u_{\alpha_1^j}\}$ and $\{v_{\alpha_1^j}\}$,
respectively,
and  show that  as $\alpha_1^j \to 0^+$
\[
\tilde B (u_{\alpha_1^j}, v_{\alpha_1^j}) \to \tilde B(u,u) = B(u,u)
\qquad {\mbox{weakly in }} \qquad  L^{2} ([0,T]; D(A)\sp) \; ,
\]
by following an approach similar to that used  in the proof of
Theorem~\ref{Global}. This finishes the proof of the Theorem.

\end{proof}

\bigskip

\section*{Acknowledgments}

This work was supported in part by the National Science Foundation.
The main part of this  research  was performed while CF and EST were visiting
scholars at the Los Alamos National Laboratory in 1997. We would like to 
thank S. Chen, C. F. Keller and L. G. Margolin for kind hospitality and
encouragement.

\end{document}